\documentclass[aps,prl,twocolumn,superscriptaddress,groupedaddress]{revtex4}

\usepackage{graphicx} 
\graphicspath{ {images/} }

\usepackage{paralist} 
\usepackage{natbib}

\begin{document}
\title{Direct observation of ballistic Andreev reflection}
\author{T.M.Klapwijk}  
\email{t.m.klapwijk@tudelft.nl}

\affiliation{Kavli Institute of NanoScience, Faculty of Applied Sciences, Delft University of Technology, Lorentzweg 1, 2628 CJ Delft, The Netherlands}

\affiliation{Laboratory for Quantum Limited Devices, Physics Department, Moscow State Pedagogical University, 29 Malaya Pirogovskaya St., Moscow, 119992, Russia}
\author{S.A.Ryabchun}
\affiliation{Laboratory for Quantum Limited Devices, Physics Department, Moscow State Pedagogical University, 29 Malaya Pirogovskaya St., Moscow, 119992, Russia}
\affiliation{Moscow Institute of Electronics and Mathematics, National Research University Higher School of Economics,
Moscow 109028, Russia}
\date{July 30, 2014}


\maketitle

An overview is presented of experiments on ballistic electrical transport in inhomogeneous superconducting systems which are controlled by the process of Andreev reflection. The initial experiments based on the coexistence of a normal phase and a superconducting phase in the intermediate state led to the concept itself. It was followed by a focus on geometrically inhomogeneous systems like point contacts, which provided a very clear manifestation of the energy- and directional dependence of the Andreev reflection process. The point contacts have in recent years evolved towards the atomic scale by using mechanical break-junctions, revealing in a very detailed way the dependence of Andreev reflection on the macroscopic phase of the superconducting state. 
In present day research the superconducting inhomogeneity is constructed by clean room technology and combines superconducting materials with, for example, low-dimensional materials and topological insulators. Alternatively the superconductor is combined  with nano-objects, such as graphene, carbon nanotubes, or semiconducting nanowires. Each of these 'inhomogeneous systems' provides a very interesting range of properties, all rooted in some manifestation of Andreev reflection.

\section{Introduction}
The 50-year-old concept of Andreev reflection\cite{Andreev1964a}, published in May 1964,  arose originally in the context of ballistic transport in \emph{inhomogeneous} crystalline materials with parts in the superconducting phase intermixed with parts in the normal phase. The difference between the electrical and the thermal conductivity, already observed in the early '50s by Mendelssohn and Olsen\cite{MendelssohnOlsen1950} and Hulm\cite{Hulm1953} was not resolved by the 1959 microscopic theory of the thermal conductivity by Bardeen et al\cite{BardeenThermal1959}. Subsequent experimental work by Zavaritskii\cite{Zavaritskii1960} in 1960 and by Str\"assler and Wyder\cite{StrasslerWyder1963} in 1963 led Andreev to the analysis of electron transport at the interface between the normal and the superconducting phase in one and the same crystal. He identified the unique process of the conversion of an electron into a hole which retraces the path of the incident electron, accompanied by the simultaneous process of a charge of 2e being carried away by the superconducting condensate. This process facilitated charge transport but it did not allow for energy transport and the observed thermal boundary resistance was a natural consequence \cite{Andreev1964a,Andreev1964b}.  

An interface between a normal metal and a superconductor is an example of an inhomogeneous superconducting system. Since the early '50s the natural framework for dealing with a position dependent superconducting order parameter was provided by the Ginzburg-Landau theory\cite{GinzburgLandau1950}. The original BCS-theory\cite{BCSLetter18-2-1957,BCSFull8-7-1957} assumed a uniform superconducting state. By developing a formulation in 1958 of the microscopic theory\cite{Gorkov1958}, which allows for spatial variations, Gorkov\cite{Gorkov1959} showed in 1959 that the Ginzburg-Landau theory can be derived from the microscopic theory. The Ginzburg-Landau theory is only valid close to the critical temperature $T_c$, whereas the difference in thermal and electrical conductivity was primarily manifest at temperatures much lower than $T_c$. A conceptual framework for inhomogeneous superconductors was needed, which included the spectral properties of the superconducting state, which is available in the original Gorkov-theory\cite{Gorkov1958}. The now commonly used Bogoliubov-De Gennes equations are a limiting case of these Gorkov equations, suitable for treating ballistic transport.            

From an experimental point of view another very important step was taken almost simultaneously in 1965 by Sharvin\cite{Sharvin1965a} by the invention of mechanically constructed metallic point contacts. This allowed the study of electrical transport between two dissimilar materials, with electrical transport governed by classically ballistic electrons. The application of this concept of ballistic transport to normal metal-superconductor contacts provided the framework, introduced by Blonder et al\cite{BTK1982}, to measure very directly the energy-dependence of the Andreev-scattering process. The Sharvin point contacts stimulated also a new approach to the description of electrical transport on the nanoscale level by using the scattering matrix approach, introduced already in 1957 by Landauer\cite{Landauer1957} and generalized and applied to phase-coherent normal transport in nanoscale objects by B\"uttiker in 1985\cite{Buttiker1985}. Rather than relying on a general theory for inhomogeneous systems it focuses on simplified experimental systems in which the phase-coherent transport problem can be split into three pieces. It selects the class of problems in which two equilibrium reservoirs can be defined, usually at a different chemical potentials or temperatures, which serve as emitters or absorbers of quantumparticles and a scattering region in which the interesting physical processes occur and which can be characterized by a scattering matrix with certain symmetry properties. 

The experimental progress in constructing nano-objects with the now universally available clean room technology has led to many experiments based on nano-objects connected to superconducting rather than normal metal reservoirs. This leads to a large variety of objects and observations in which the challenge is to discover new phenomena and at the same time establish through transport experiments what one has actually made in the clean room. In some cases the general theory of inhomogeneous nonequilibrium superconductivity is used to interpret these specific cases.  At the same time the perceived unique nature of these nano-objects has led to an application of the scattering matrix approach in which the superconducting contacts serve as equilibrium reservoirs, which communicate with the scattering region through the Andreev reflection process. An experimental challenge is to determine which framework is appropriate for the actual nano-objects emerging from the clean room and where theoretical innovation is needed.

In the following I have made an attempt to summarize the developments in the subject over the past 50 years. The focus is on experimental observations, which provide a direct demonstration related to ballistic Andreev reflection. The main attention is paid to the demonstration of the reversal of direction, as well as of the charge, and the spectroscopically important dependence on energy.  Furthermore, a third important aspect is the dependence on the macroscopic quantum phase, which manifests itself when more than one superconductor are used. It  leads to the concept of Andreev bound states, which carry the Josephson current. Since the field has become large, a further selection was applied by focusing on experiments which are sufficiently well defined that a quantitative description turned out to be possible. Needless to say that many experiments are not included, in particular those in which diffusive scattering is the dominant ingredient. The Section-headings give an indication of the subject. They are supplemented with the dates in which, in my view, the most significant developments for this subject took place.       

\section{Inhomogeneous superconductivity close to $T_c$: ~1950-1957}
After the discovery of superconductivity by observing \emph{zero resistance}  by Kamerlingh Onnes in 1911 it took until 1933 for a second fundamental property to be identified by Meissner and Ochsenfeld called \emph{perfect diamagnetism}. An early explanation was provided by Fritz and Heinz London in 1935 by a modification of the Maxwell equations inside a superconducting material. It was known that these properties are very nicely observed in pure crystals of tin, aluminium and mercury. However, it was also known that many superconducting alloys did not obey these basic relations. In particular, perfect diamagnetism was not observed although the material provided zero resistance. Apparently, magnetic flux was not completely excluded and the magnetization curve was not reversible but showed clearly hysteretic effects. The first theory capable of handling inhomogeneous systems was the Ginzburg-Landau theory, introduced in 1950. It was used by Abrikosov in 1957 to analyze what would happen with a superconductor if the magnetic penetration depth known from the London-theory, $\lambda_L$ would exceed another characteristic length $\xi$ called now the Ginzburg-Landau coherence length. 

By minimizing the expression for the free energy for a volume in which the order parameter may vary with position one finds the two celebrated Ginzburg-Landau expressions: 

\begin{equation}
\label{GLdiffequation}
\frac{1}{2m^*}\left [-i\hbar \nabla-\frac{e^*{A}}{c}\right ]^2 \psi+\alpha \psi + \beta |\psi|^2\psi=0
\end{equation}
and 
\begin{equation}
\label{GLsupercurrent}
j=\frac{e^*\hbar}{2im^*}(\psi^*\nabla \psi-\psi\nabla \psi^*)-\frac{{e^*}^2}{m^*c}\psi^*\psi A
\end{equation}
These two equations make it possible to calculate the order parameter as a function of position in the presence of a magnetic field, including the distribution of the current. The magnetic field $H$ is the locally present field strength. And of course it is assumed that the order parameter $\psi$ is complex with a phase $\phi$, which also may be position-dependent.

The most ideal inhomogeneous system is one in which we have a clearly defined boundary between a piece of atomic matter in the \emph{superconducting state} and a piece of the \emph{same} atomic matter in the \emph{normal state}. In such a system no barrier would be encountered for normal electronic transport, because the material is uniform in its atomic arrangement. Obviously, this is not the case in the many nano-devices studied today, which consist of different materials with different atomic arrangements. This non-uniformity in the atomic sense, which goes beyond the superconducting properties, is an experimental nuisance. It contributes however strongly to the interplay between elastic and Andreev scattering. Below I sketch two cases in which this complexity is absent.

\subsection{Inhomogeneous system due to an applied magnetic field: intermediate state.}

For Type II superconductors, discovered by Abrikosov,  $\lambda$ is much larger than $\xi$ and quantized vortices are the dominant inhomogeneous state, with their own interesting microscopic properties. In the other limit, $\lambda << \xi$,  in the presence of a magnetic field the material breaks up in lamellae of alternating superconducting and normal phase slabs. One of the attractive features of this intermediate state in Type I superconductors is that it provides a system with uncompromised interfaces between a normal state and a superconducting state. In one and the same material with the same atomic constituents one has a domain with electrons in the normal state and a domain with electrons in the superconducting state. The price one pays is that the normal state is in the presence of a magnetic field. However, the normal state is hardly effected by the presence of this magnetic field.
  
For a material such as tin or aluminium a single crystal can be grown with an elastic scattering length in the order of millimeters. The ratio of the resistance at room temperature compared to the one at low temperatures can be in the range of several 10000s. The crystals can have a high degree of purity with an impurity resistance very low compared to the resistance due to electron-phonon scattering. These crystals have been used extensively to study the transport properties. The electrical resistance in this intermediate state is very well understood as being due to the resistivity of the normal state multiplied by the thickness of the normal slabs and their number. The difficulty was that the thermal resistance did not behave in the same way. It appeared as if there was a thermal boundary resistance present, which increased by lowering the temperature. This difference was the starting point for the concept of Andreev reflections.  
       
\begin{figure}[t]
\includegraphics[width=9cm]{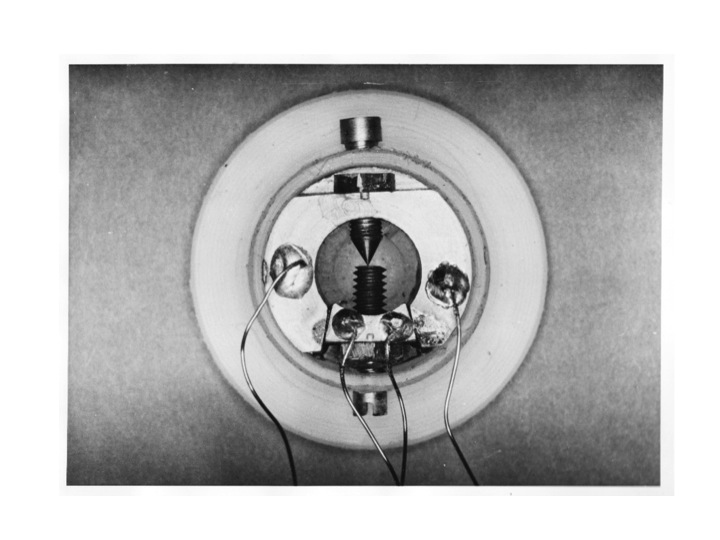}
\centering
\caption{\label{fig:Pointcontact}Superconducting point contact made of a pointed niobium screw touching a niobium anvil, both in a yoke of superconducting material, separated by a thin layer of glass with the same thermal expansion coefficient as the niobium. From the perspective of the electrons the actual contact is formed by a metallic path punching through the surface oxide.}
\end{figure}

\subsection{Constriction-type inhomogeneity}
Another example of an inhomogeneous system is a constriction-type Josephson weak link\cite{Likharev1979}. Let's consider two massive volumes of superconductor, which are only linked to each other at one point by a short and narrow piece of the same superconductor (see Fig.~\ref{fig:Pointcontact}). In the absence of a magnetic field and a current the superconducting order parameter $\psi$ is everywhere the same. If a current is applied the current-density will be low in the banks of the point contact and high in the neck, where a strong gradient of $\psi$ will be present. Aslamazov and Larkin\cite{AL1969} have analyzed this case, starting with the observation that in Eq.~\ref{GLdiffequation} the dominant term is the 2nd derivative term: 

\begin{equation}
\label{LaplaceEquation}
\nabla^2\psi=0
\end{equation}    
which has to be solved together with Eq.~\ref{GLsupercurrent}. The inhomogeneity is in this case due to the different cross-sections which enforce a strong difference in current-density.  

The current can be expressed using the second GL equation (Eq.\ref{GLsupercurrent}) leading to:

\begin{equation}
J_s=C |\psi_1| |\psi_2|\sin(\phi_1-\phi_2)
\end{equation}

This simple derivation has shown in an elegant way that the characteristic $\sin\phi$ dependence of the Josephson-effect emerges quite generally, close to $T_c$, for both dirty and clean superconductors. The important assumption is that two equilibrium reservoirs are connected by a weak link, which allows one to reduce the problem to solving Eq.\ref{LaplaceEquation}, in this case under the assumption that the Ginzburg-Landau equations can be applied, valid for small values of the order parameter ($\Delta<<k_B T$). It is customary to assume that in the presence of a voltage the supercurrent has a parallel current given by Ohm's law, $I_n=V/R$, with $R$ a voltage independent resistor, without any information about the microscopic superconducting properties. In subsequent research it has become clear that the voltage-dependence of the 'normal' transport, \emph {i.e.} the non-linearity of the resistor,  is a rich source of information about the microscopic properties.  In experiments it is unavoidable, given the way the point contacts are made that there is a possibility of enhanced elastic scattering at the constriction itself, unlike in the previous case of the intermediate state.         

\section{Inhomogeneous systems far below $T_c$:~1963-1966}
The original microscopic theory of superconductivity considers a uniform system. Within this framework the electronic thermal conductance was calculated by Bardeen, Rickayzen and Tewordt\cite{BardeenThermal1959}, showing an exponential decay of the electronic contribution to thermal conductivity, in line with the reduction of the quasi-particle density. Given the examples above the challenge is to deal also with inhomogeneous superconducting systems in the limit of $\Delta>>k_B T$. In the opposite limit of $\Delta<<k_B T$, where the Ginzburg-Landau equations can be derived from the microscopic theory, it was shown that the very useful explicit expressions for $\xi$ and $\lambda$ in both the clean and dirty limit can be derived (such as for example given by Saint-James et al\cite{SaintJames1969})  and led to the identification of $e^*=2e$, $n_s=n/2$. The numerical coefficients are fixed using the convention that $m^*=m$, the free electron mass. Nevertheless, since the Ginzburg-Landau equations are limited to the range close to $T_c$, only properties which depend on the value of the order parameter and its phase can be handled.   

In the experiments it was found that in very good atomically uniform crystals of superconductors, such as mercury\cite{Hulm1953} or indium\cite{StrasslerWyder1963} with, at low magnetic fields, a good Meissner state, $\vec B=0$, upon application of a magnetic field the domains appeared, which were in the normal state (N) interleaved by domains which were in the superconducting state (S). The crystals studied had a mean free path for elastic scattering in the order of $0.5$ mm, whereas the thicknesses of the N and S layers were inferred to be in the $0.02$ mm range. In other words the transport at the NS interfaces could definitely be considered as ballistic. 

In 1964 Andreev\cite{Andreev1964a} uses the Gorkov equations\cite{Gorkov1958}, applied to a system, without impurity scattering, which contains a more or less sharp boundary between a normal phase and a superconducting phase. He finds the conversion of an electron to a hole with a probability which depends on the energy relative to the energy gap $\Delta$ of the superconducting state. He proceeds by calculating the thermal flux, across the boundary, and compares the result with the data obtained by Zavaritskii. In passing he points out that the path of the electron and the hole have a unique element to it:  \emph{We note the following curious feature. Usually when particles are reflected, only the component of the velocity normal to the boundary changes sign. The projection of the velocity on the plane of the boundary remains unchanged. In our case all three components of the velocity change sign.}    
It means that the reflection process is dependent on the energy, it inverts the charge and it leads to a reversal of all velocity-directions.  

Although Andreev based his analysis on the Gorkov-equations, the most common approach to discuss the process of Andreev reflection is now by using the Bogoliubov-De Gennes equations. However, Bogoliubov and De Gennes never wrote a paper together and it is therefore worthwile to provide some indication on how these names came together. The emergence of the Bogoliubov formulation of the theory of superconductivity together with the construction of the Gorkov-theory is described by Gorkov\cite{Gorkov2011}. Around 1963 Pierre Gilles de Gennes applied the Bogoliubov-transformation to a position-dependent eigenfunction. He defines:   
\begin{equation}
\label{BogolTransform}
\psi(\vec r \uparrow)=\sum_n[\gamma_{n\uparrow}u_n(\vec r)-\gamma^*_{n\downarrow}v^*_n(\vec r)]
\end{equation}       
which represents the annihilation operator for a position eigenfunction, with $u$ and $v$ also position-dependent  eigenfunctions to be determined from the effective Hamiltonian with $\Delta(\vec r)$ to be found self-consistently from:
\begin{equation}
\label{BCSgapequation}
\Delta(\vec r)=V<\Psi(\vec r \uparrow)\Psi(\vec r \downarrow)>
=V\sum_n v^*_n(\vec r)u_n(\vec r)[1-2f_n]
\end{equation}
From this starting point De Gennes derives the set of coupled equations, which are now called Bogoliubov-De Gennes equations. They appear for the first time in print in 1963 in De Gennes and Saint-James\cite{DeGennesSaintJames1963}. It is stated that for a normal metal film on a superconductor the one-particle excitation energies are the eigenvalues of: 
\begin{eqnarray}
\label{BDGeq}
\nonumber
Eu & = & \left(-\frac{\hbar^2}{2m}\nabla^2-E_F\right)u+\Delta v \\
Ev & = & \left(\frac{\hbar^2}{2m}\nabla^2+E_F\right)v+\Delta u
\end{eqnarray}
The "pair-potential" $\Delta$ is defined as 
\begin{equation}
\label{Gap}
\Delta(\vec r)=g(\vec r)<\psi(\vec r)\psi(\vec r)>
\end{equation}
with $g(\vec r)$ the local value of the electron-electron coupling constant and the $\psi(\vec r)$ the usual one electron operators.

This set of equations (Eqs.\ref{BDGeq}), which obviously look like a set of Schr\"odinger equations coupled by the parameter $\Delta$ are called the Bogoliubov-De Gennes equations. To the best of my knowledge the assignment of these equations to these two authors together, and not for example to De Gennes and Saint James, is for the first time done in print in the paper by Kulik\cite{Kulik1969} on the supercurrent in a SNS junction. Historically it is clear that the origin can be found in the self-consistent field method for the BCS theory of Bogoliubov\cite{Bogoliubov1958c}, which was originally published in ZhETF \cite{Bogoliubov1958a} and Il Nuovo Cimento\cite{Bogoliubov1958b}. Kulik refers to another paper of Bogoliubov\cite{Bogoliubov1959}, which deals with general aspects of the self-consistent field method. 

The actual derivation of the BdG equations is given by Saint-James in Ref.\cite{SaintJames1964} in an Appendix, while referring to Lecture notes on the subject of De Gennes, dated 1963-1964, which were later published by De Gennes\cite{DeGennes1966} in 1966. Ironically, in a 1964 paper on the excitations in a vortex core Caroli, De Gennes and Matricon\cite{Caroli1964}, refer to the set of equations by simply citing Bogoliubov \emph{et al}\cite{Bogoliubov1958a}. Unfortunately, this hides a major accomplishment by De Gennes, which is the generalization of the Bogoliubov $(u,v)$ transformation to the case of inhomogeneous systems. Shirkov\cite{Shirkov2009}, a former collaborator of Bogoliubov\cite{Bogoliubov1958a}, calls it the Bogoliubov-De Gennes transformation (Eq.\ref{BogolTransform}) that can be written in terms of coordinate-dependent $u(\vec r), v(\vec r)$ wave functions of electrons in the superconducting phase. The conclusion is that the major step forward of De Gennes was the generalization of the Bogoliubov transformation to position-dependent wave-functions for the quasiparticles through which he opened the door to treat ballistic inhomogeneous problems in superconductivity. It is therefore historically understandable to call the set of equations (Eq.\ref{BDGeq}) the Bogoliubov-De Gennes equations. At the same time it is clear that the Gorkov equations are more general and can be used as starting point for treating also the cases with diffusive scattering and nonequilibrium problems.     

Meanwhile, in the course of history, the significance of the contribution of Saint James might have become underexposed.  Interestingly, in the 1964-paper by Saint-James one finds already a glimpse of the phenomenon of what we now call Andreev reflection. He publishes the more extensive calculation of the 1963 paper with De Gennes\cite{DeGennesSaintJames1963}, using the Bogoliubov-De Gennes equations, for a normal metal of thickness $a$ on a superconductor with the interface located at $x=0$  to determine the excitation spectrum. At the end of this calculation he writes in the French Journal de Physique: \emph{What is the origin of the result? An electron travels through the region (N), and penetrates in (S), where it creates an electron-hole pair. The two electrons combine to form a Cooper-pair, leading to a hole which travels back inside (N), after which it reflects at the opposite surface of (N) at $x=-a$ and the cycle will repeat. The total duration of the cycle is $4a/{v_F} cos\theta$}
\footnote{Quelle est l' origine physique de ce r\'esultat? Un \'electron s'avance dans la r\'egion (N), p\'en\'etre dans (S) o\`u  il cr\'ee une paire \'electron-trou. Les deux \'electrons se combinent pour former une paire de Cooper, tandis que le trou repasse dans (N), se r\'efl\'echit sur la surface $x=-a$ et revient dans (S) o\`u il d\'etruit une paire de Cooper. Un \'electron appara\^it de nouveau, repasse dans (N), se r\'efl\'echit sur la surface et le cycle recommence. La dur\'ee totale de ce cycle est: $4a/{v_F} cos\theta$.} with $\theta$ a measure of the energy. The factor of $4$ indicates that the slab needs to be traversed two times to provide interference to be contrasted with $2$ for normal reflection.  Based on this article Deutscher\cite{Deutscher2005} has recently argued that the phenomenon of Andreev reflection should be called Andreev-Saint-James reflection to do justice to the historical record. In my view the unique nature of the process of Andreev reflection is the reversal of \emph{all} velocity-components, the unfamiliar process called retro-reflection, which is fully recognized and understood for the first time in the original Andreev paper\cite{Andreev1964a}. Therefore I believe it is justified to continue to speak about the concept of Andreev reflection, meaning the reversal of all velocity components and the charge.

The framework of the Bogoliubov-De Gennes equations (Eqs.\ref{BDGeq} and \ref{Gap}) allows for a description of a non-uniform superconducting state in many selected cases of current interest. The parameter $V$ in Eq.\ref{BCSgapequation} takes care of the attractive interaction leading to superconductivity. The quantity $\Delta$ can be present anywhere, expressing what is called the proximity-effect.  The Bogoliubov-De Gennes equations have been used extensively to determine the excitation spectrum for materials in which, under certain conditions the normal and superconducting phase coexist. Examples are the excitations in the core of a vortex\cite{DeGennesSaintJames1963}, the excitations in the normal domain of a Type I superconductor in the normal state\cite{Andreev1966} and the excitations in an SNS type Josephson-junction by Kulik\cite{Kulik1969}. In the latter two cases the calculation is usually carried out for a one-dimensional model. 

Most experiments were carried out on high purity, well-annealed single crystals of tin, indium, mercury or lead. In these samples the elastic mean free path reaches easily a size approaching a millimeter. Therefore it was natural to ignore elastic scattering and to treat the wave-functions as plane waves. A direct measurement of the excitation spectrum had to wait, in all 3 cases, for the arrival of nanolithography and scanning probe techniques. On the other hand the concept of Andreev reflection nicely explained the observed difference between the electrical and thermal conduction at NS interfaces. The remaining question is what direct experimental evidence has been accumulated to test the theoretical ideas in a qualitative and quantitative way. How would one get experimentally access to a well-defined normal metal-superconductor interface, for which one can study qualitatively and quantitatively the process of Andreev reflection itself?  

\section{Ballistic transport and electron focusing: ~1966, 1974}
\begin{figure}[t]
\includegraphics[width=8cm]{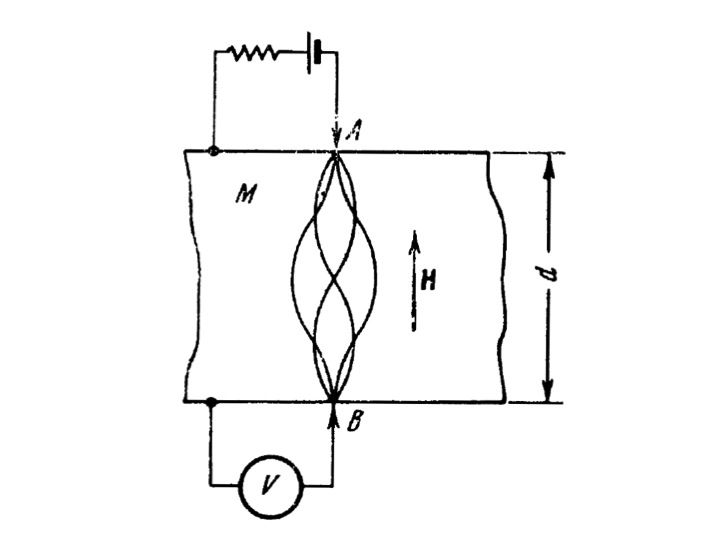}
\centering
\caption{\label{fig:Sharvin1969b}Ballistic trajectories in a pure crystal following paths controlled by the Fermi surface. Pointcontact A acts as the emitter and pointcontact B  as the collector. A magnetic field is applied in the direction of the electron flow. (Picture taken from Sharvin and Fisher\cite{Sharvin1965b}.}
\end{figure}

In hindsight, in order to be able to study and exploit the phenomenon of Andreev reflection in its full potential one needs a source of quasi-particle waves, a collector and a medium through which their properties are manifest. One of the first steps along this path was set by Sharvin\cite{Sharvin1965a}, who introduced a new technique to study Fermi-surfaces by putting a sharp metallic needle on a bulk single crystal of a metal as a source and a second one at the opposite side as a collector (Fig.\ref{fig:Sharvin1969b}). The electrons would follow paths along the Fermi surface and the trajectory between source and collector could be influenced by a magnetic field. In his analysis he treats the point contact as ballistic \emph{i.e.} with a size small compared to the elastic mean free path in the material of the needle as well as of the crystal. From this assumption he infers that the current is the difference between electrons coming from one reservoir at voltage V, while the other reservoir, kept at ground, sends electrons in the other direction. The 'Sharvin' resistance is then given by:
\begin{equation}
R=\frac{p}{e^2D^2N}
\end{equation}
with $D$ the diameter of the hole forming the point contact, $p$ the Fermi momentum, and $N$ the electron density.
A first observation was carried out by Sharvin and Fisher\cite{Sharvin1965b} and in more detail by Sharvin and Bogatina\cite{Sharvin1969}. Since the mean free path is much larger than the diameter of the orifice $D$ the resistance is not the familar backscattering resistance inside the constriction, but rather the geometrical restriction on possible conduction channels.   

This pioneering work with point contacts to understand fundamental transport processes, led to two new types of experiments. Yanson\cite{Yanson1974} recognized in 1974 that the concept of ballistic transport through an orifice as introduced by Sharvin was useful to understand 'failed' tunnel junctions. By applying a large voltage to a tunnel barrier one obtains one or more leaky pathways, which can be analyzed as a Sharvin contact. He went one step further and pointed out that the current-voltage characteristic might contain non-linearities due to a backscattering current resulting from phonon-excitation, which should reveal the electron-phonon interaction spectrum. It was known that the latter was measurable with superconducting tunnel junctions. This new pointcontact technique allowed for a measurement of the electron-phonon interaction in normal metals. The success was demonstrated for copper by Yanson and Shalov\cite{Yanson1976}. It inspired a group in Nijmegen in the Netherlands, led by Peter Wyder (who had  interest in point contacts for far-infrared detection; see the following Section) to apply the same reasoning to the point contacts as used by Sharvin. First results of this technique applied to copper, silver and gold were published by Jansen et al\cite{Jansen1977} and the method was popularized by a publication in Science by the same authors\cite{Jansen1978}.  

A second development was introduced also in 1974 by Tsoi\cite{Tsoi1974,Tsoi1975}. He modified the technique of Sharvin to follow the paths of the electrons by putting the source and the collector on the same side of the crystal. By using a transverse magnetic field he was able to tune the cyclotron orbits in such a way that for certain specific strengths of the magnetic field the emitted electrons would reach the collector preferentially, also reflecting the Fermi-surface properties. This technique was also adopted by the Nijmegen group, leading to a collaboration between the group at Chernogolovka and at Nijmegen\cite{TsoiWyder1979}. 

This work with point contacts has laid the groundwork for an understanding of transport in terms of classical ballistic trajectories. It meant a concept of electronic transport in which two equilibrium reservoirs are connected through a small orifice with a radius $a$, which has a net resistance of:
\begin{equation}
R=\frac{4\rho l}{3\pi a^2}
\end{equation}
with $\rho l$, the so-called $\rho l$-product given by the free electron values: ${mv_F}/{ne^2}$.   
Electrons passing through the orifice are absorbed by the reservoir where they equilibrate and conversely, the reservoirs act as sources of equilibrium electrons.

\section{Josephson point contacts:~1966, 1979}

In parallel to the research on the use of normal metal point contacts there was quite a bit of research of a more applied nature on superconducting point contacts such as the one shown in Fig.\ref{fig:Pointcontact}. Superconducting point contacts have been extensively used in early developments of SQUID magnetometers and to demonstrate the response to radiation known as Shapiro steps. Undoubtedly, one of the beautiful aspects of the Josephson effect is that it is a macroscopic quantum phenomenon, which can occur in any kind of weak links, between two superconductors. Whatever the type of the weak link,  if the coupling is not too weak to be disrupted by thermal or quantum noise, any material put between the two superconductors, even vacuum, will offer a manifestation of the basic characteristics of the Josephson effect.  After the initial observation in a tunnel junction by Rowell and Anderson, it was quickly followed by a demonstration of the ac Josephson effect in a superconducting microbridge, sometimes called an Anderson Dayem\cite{DayemAnderson1964} bridge. Zimmerman and Silver\cite{ZimmermanSilver1966} introduced in 1966 a DC SQUID based on two mechanically made point contact diodes, very much like the Sharvin point contacts. The technique of using point contacts was quickly taken up by researchers interested in an excellent coupling to microwave radiation. Dayem and Grimes\cite{DayemGrimes1966} studied the emitted radiation of a point contact biased at a certain voltage.  Levinstein and Kunzler\cite{LevinsteinKunzler1966} showed that a point contact made it possible to obtain current-voltage characteristics which evolve from a tunneling curve to a typical point contact I,V curve, of which at that time the nature was not yet fully understood. Grimes, Richards and Shapiro\cite{GrimesRichardsShapiro1966,GrimesRichardsShapiro1968} turned the point contact into a detector of far-infrared radiation. The technical details of their apparaturs have been described by Contaldo\cite{Contaldo1967}.

\begin{figure}[t]
\includegraphics[width=9cm]{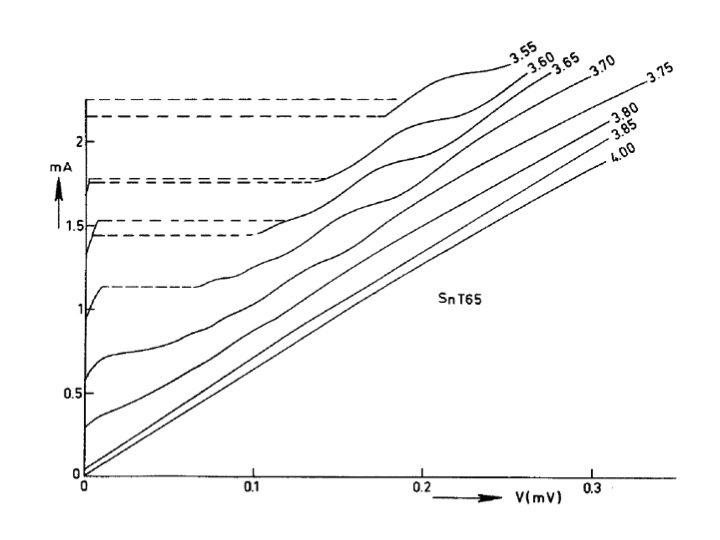}
\centering
\caption{\label{fig:microbridge}Typical current-voltage characteristics for variable-thickness microbridges, made of superconducting tin. These curves clearly show all the salient deviations from the RSJ-model. A resistive state at low voltages with a slope much smaller than the normal state resistance, subharmonic gap structure and an excess current. The data are all taken close to $T_c$ because at lower temperatures thermal hysteresis dominates and masks the interesting physics. From Klapwijk \emph{et al.}\cite{Klapwijk1977}.}
\end{figure}

Meanwhile, the scientific concepts around the point contacts of  Zimmerman and Silver were quite different from those of Sharvin. For Zimmerman and Silver the microscopy of the Sharvin point contact appeared to be completely absent. The emphasis was on the electromagnetic performance. Since all of the point contacts, as well as the microbridges,  had a low normal state impedance it was understood by Stewart\cite{Stewart1968} and McCumber\cite{McCumber1968} that the most appropriate engineering model was that of the resistively-shunted model (RSJ-model), which could be shunted by a capacitor and is therefore often called the RSJC-model. This RSJC-model treated the point contact as a Josephson element characterized by the celebrated Josephson equations and shunted by a capacitor. This model made it possible to understand the dominant difference between a Josephson tunnel junction and a low-capacitance current-biased point contact or microbridge. It also made it possible to identify at which level of capacitance hysteresis would appear in the IV-curve. The Stewart-McCumber model became the paradigm for all research in which a microscopic understanding was not needed or not looked for. In reality there were very many deviations (see for example Figs.~\ref{fig:microbridge} and \ref{fig:weitz3}), which were temporarily ignored. It is still the dominant model for experiments in which the Josephson-junction functions as a building block for macroscopic quantum tunneling.           

In the former Soviet Union research on point contacts and microbridges aimed at the interaction with high frequency radiation was picked up at several laboratories of the Academy of Sciences. Early work was found at the Institute of Physics Problems by Khaikin and Krasnopolin\cite{KhaikinKrasnopolin1966} in 1966. A strong program led by Vystavkin and Gubankov emerged at the Institute of Radioengineering and Radioelectronics of the Academy of Sciences around 1970. Working at this laboratory Volkov and Nad'\cite{VolkovNad1970} published experimental data and a theoretical analysis of Shapiro steps observed in niobium point contacts, using the theory presented by Aslamazov and Larkin\cite{AL1969}, which in essence is the Stewart-McCumber model. The interest in constriction-type Josephson junctions is clear from the 1974 review paper by Vystavkin \emph{et al}\cite{Vystavkin1974} in which research at the IREE is presented together with work from Likharev at Moscow State University. Both point contacts as well as superconducting microbridges were developed and studied. The majority of the work was focused on the Josephson effect and interpreted within the framework of the RSJC model. However, as in many other groups significant deviations form the RSJC-model were observed. In a number of cases the solution was sought within the lumped circuit nature of the RSJC model. Although there was a strong drive towards using Josephson junctions for practical applications a number of people stepped out of that mode and focused on an improved microscopic understanding. In reality a new microscopic theoretical framework, largely absent in the well-known review of Likharev\cite{Likharev1979} from 1979, was needed  to deal with inhomogeneous problems in nonequilibrium superconductivity. 

\section{Diffusive nonequilibrium theory: ~1968-1979}
The Bogoliubov-De Gennes equations had emerged as suitable in dealing with inhomogeneous problems in systems with little or no impurity scattering. However, the superconducting devices that were of interest for practical applications, the point contacts and microbridges, were made of materials which had significant impurity scattering. The nature of the contact of the point contact was not very well known, but given the crude way of making them elastic scattering inside the contact was to be expected. Microbridges were made from vacuum-condensed thin films and made with diamond knife technology or primitive lithography. In order to describe these inhomogeneous superconducting systems, including impurity scattering, an appropriate theoretical framework was urgently needed.      

In the late '60s and early '70s the quasiclassical theory for inhomogeneous and nonequilibrium superconductivity was developed. It started with the Gorkov-theory\cite{Gorkov1958}, with the subsequent developments primarily in the former USSR with significant contributions in Germany. A general introduction is available through the textbook by Kopnin\cite{Kopnin2001}. An overview convenient in the context of mesoscopic systems is provided by Belzig et al\cite{Belzig1999}. For the topic addressed here the main message is that this framework provides a microscopic theory for inhomogeneous systems, which is valid for all temperatures and which is also suitable for nonequilibrium systems. So the theory is particularly well suited, although not always easily tractable, for superconducting constrictions such as microbridges and point contacts, including cases when the scattering is diffusive. The theory is, in principle, also well suited to deal with the large variety of modern hybrid devices in which nano-objects are coupled to superconducting electrodes. 

The starting point is the field-theoretical description of superconductivity introduced by Gorkov\cite{Gorkov1958}, which has evolved into the quasiclassical theory by removing the rapid oscillations on the scale of the Fermi wavelength by Eilenberger\cite{Eilenberger1968}, and Larkin and Ovchinnikov\cite{LarkinOvchinnikov1968}. In order to deal with finite temperatures the Matsubara\cite{Matsubara1955}-frequencies and Keldysh\cite{Keldysh1964} techniques are used. A distinction can be made between clean and dirty systems resulting for dirty systems in the theory for nonequilibrium inhomogeneous superconductivity problem of Schmid and Sch\"on\cite{SchmidSchon1975}, and Larkin and Ovchinnikov\cite{LarkinOvchinnikov1975,LarkinOvchinnikov1977}.
    
\begin{figure}[t]
\includegraphics[width=9cm]{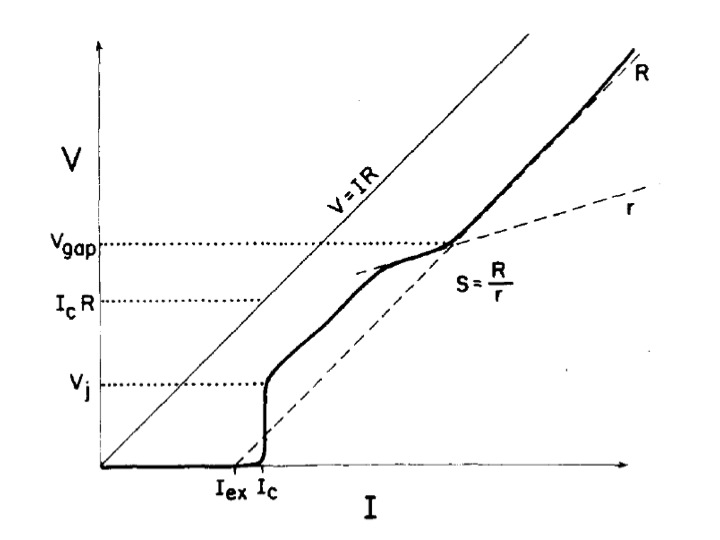}
\centering
\caption{\label{fig:weitz3}Current-voltage characteristic for a ballistic niobium point contact with ideal equilibrium reservoirs (taken from Weitz et al\cite{Weitz1978}). It clearly shows the excess current beyond $V_{gap}$ as well as the overall deviations from the resistively shunted junction model. Similar I,V curves have been obtained for niobium tunnel-junctions with a 'leaky' silicon barrier.}
\end{figure}

The approximations made over this 10 year period have been very helpful in making the theory usable for the study of Josephson devices such as point  contacts and microbridges. It was applied to a number of outstanding problems in the field of superconducting contacts. In contrast to tunnel junctions of which the quasiparticle current branch, the Giaever tunneling, was understood even prior to the discovery of the Josephson effect, the voltage-carrying state of superconducting point contacts contained a number of poorly understood phenomena (see for example Fig.~\ref{fig:microbridge}). First, when exceeding the critical current a steep rise in current, at finite voltage, is observed up to a few microvolts after which point the voltage increases much more rapidly and often discontinuously. This 'knee-structure' or 'foot-structure',  depending on whether one plots the voltage or the current horizontally,  was observed in various laboratories and violated the elementary RSJ-model. In addition, upon further increase of the current the well-known subharmonic gapstructure, features in the IV-curve at $2\Delta/n$ are observed, followed by a so-called excess current beyond $2\Delta$, a shifted asymptote suggesting a capacity to carry more current at the same voltage than in the normal state. A very clear example of the excess current can be found in Weitz et al\cite{Weitz1978} and reproduced in Fig.\ref{fig:weitz3}. These phenomena were universally observed in all constriction-type superconducting contacts, such as the microbridges, 'pinholes' in tunnel junction barriers and in point contacts. In order to cover a large enough range of temperatures and voltages an important requirements was the use of large reservoirs to maintain thermal equilibrium in the contacts, which for microbridges led to the use of variable-thickness bridges.   

The first item, the 'knee-structure', was addressed by Golub\cite{Golub1976}, and Aslamazov and Larkin\cite{AL1976}, followed by an improved analysis by Artemenko, Volkov and Zaitsev\cite{AVZ1978b, AVZ1979ASC} and in a more accessible way by Schmid, Sch\"on and Tinkham\cite{Schmid1980}. The essential interpretation is that under the voltage bias the density of states in the neck of the constriction oscillates rapidly at the Josephson frequency. The current carried by this rapidly changing density of states behaves differently for energies $E<\Delta$ from that for energies $E>\Delta$. The first part can upon changing in time only be populated by quasiparticle relaxation due to inelastic processes. The 2nd part can easily equilibrate by diffusion to the equilibrium banks. The models assume a short one-dimensional diffusive superconducting wire connected to massive equilibrium reservoirs of the same superconducting material.

The third item that was addressed was the 'excess current'.  Artemenko \emph{et al}~\cite{AVZ1979ASC, AVZ1978a} showed in 1978-1979, that this excess current was not related to the Josephson effect, in other words unrelated to the physics contained in the RSJC-model, but was part of the static quasiparticle current through the constriction. A quite striking breakthrough came by a comparison of experimental results between S-c-S contacts with S-c-N contacts in research reported by Gubankov \emph{et al}~\cite{Gubankov1979}. It led to the decisive article of Artemenko \emph{et al}~\cite{AVZ1979} in which the theoretical results for the excess current in both S-c-S and S-c-N contacts were presented. In this paper the authors write: \emph {'Therefore, particles with energies $|E|<\Delta$ contribute to the current in the bridge. Naturally, the gap in the S-region does not prevent the charge transfer by the electrons with the energy $|E|<\Delta$. The current transferred by these particles converts into the pair current in the S-region. Note that the analogous process (called the Andreev's reflection) takes place in a pure metal when electrons pass through the ideal S-N interface.'} The message that underneath the heavy mathematical formalism the much more transparent concept of Andreev reflection was hidden was brought by Michael Tinkham to his PhD students and one of his post-docs (the present author) from a visit to Moscow in 1978. Without this explicitly made connection with the concept of Andreev reflection it would have been much more difficult to appreciate the major step forward in understanding constriction type superconducting devices. It proved that a microscopic analysis was needed  to understand the voltage-carrying state not only of tunnel junctions but also of point contact devices and that the RSJ-model was misleadingly lacking relevant microscopic input. With the word Andreev reflection for the IV-curves of Josephson point contacts on the table the conceptual framework for constriction-type Josephson junctions had to turn from phenomenological to microscopic. (The concept of Andreev reflection appeared already in the work of Artemenko and Volkov\cite{Artemenko1975} for the electrical resistance in the intermediate state based on the kinetic equations for clean superconductors proposed by Aronov and Gurevich\cite{Aronov1974}. So they were conceptually very well prepared.)

\section{Directional, charge and energy dependence of ballistic Andreev reflection: ~1980-1984}

The theory of Artemenko \emph{et al.} was based on diffusive superconductors in which the elastic mean free path is much shorter than the BCS coherence length and also than the size of the constriction. The concept of Andreev reflection was much more tailored to the picture of plane waves emanating from reservoirs analogous to the ideas of the Sharvin point contact and the subsequent implementation by Yanson and Jansen \emph{et al.} for electron-phonon spectroscopy and by Tsoi \emph{et al.} for electron focusing. After the insightful remarks about the relevance of Andreev reflection the natural starting point to take for a point contact and for microbridges was the idea of a ballistic point contact, analogous to the flow resistance of an orifice in the Knudsen gas limit. It sets the starting point for an interpretation of the second item mentioned in the previous paragraph of the subharmonic gap structure by Klapwijk, Blonder and Tinkham~\cite{Klapwijk1981}. By allowing for energy-conserving multiple Andreev processes it became immediately plausible that the subharmonic gap structure had to be understood in the same framework as the excess current, and that it was a different form of quasi-particle current flow, analogous to Giaever-tunneling but now including higher order processes. (This was quite different from the starting point for the Josephson current known from the ballistic model of Kulik\cite{Kulik1969} and Bardeen and Johnson\cite{BardeenJohnson1972}, because the process was considered to be not due to the Josephson-effect.) A description was found on the basis of the trajectory method and presented as an invited talk, resulting from rumors that we had something new to tell,  at the Low Temperature conference in Los Angeles (18-23 August 1981) and published in the proceedings\cite{Klapwijk1981}. It contained the description of the IV-curve of a NS point contacts subject to both Andreev reflection and normal reflection located at the neck of the constriction:
\begin{equation}
\label{Andreev_A}
I=\frac{1}{eR_n}\int dE[1+A(E)-B(E)][f(E-eV)-f(E)]
\end{equation}
The function $A(E)$ is called the Andreev reflection-coefficient. For $E<\Delta$ it should ideally be $1$ reflecting perfect electron-hole conversion, which would unavoidably occur at a sharp interface between the superconducting phase and the normal phase in an atomically uniform material, such as in the intermediate state:
\begin{equation}
\label{E<}
A(E<\Delta)=\frac{\Delta^2}{E^2+(\Delta^2-E^2)(1+2Z^2)^2} 
\end{equation}

\begin{figure}[t]
\includegraphics[width=8cm]{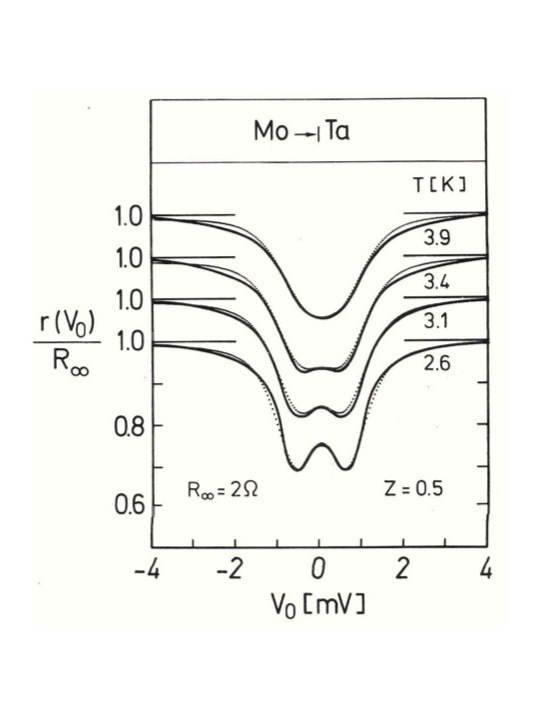}
\centering
\caption{\label{fig:Voss1984}One of the first point contact experiments showing the energy-dependent Andreev reflection coefficient weighted by the distribution functions. The fits are for the same set of parameters, only varying the temperature.  From Voss\cite{Voss1984}.}
\end{figure}

The function $B(E)=1-A(E)$ for $E<\Delta$ is the elastic back-scattering, which would be present for any elastic scattering process at the interface. It is parametrized by the parameter $Z$, which is connected to the normal state transmission ocefficient $T$ by $T=1/{(1+Z^2)}$. For $Z=0$ indeed $A=1$ and $B=0$. A similar expression controls $A$ and $B$ for energies above the gap: $E>\Delta$. Of course for high values of $E$ the Andreev reflection goes to zero but up to about $3\Delta$ there is still a significant contribution. In relation to practical experiments an important aspect is the sensitivity to $Z$ or the normal state transmission coefficient. For $Z=1$, which reflects a transmission probability of 0.5 (one would normally call it a very high transmission) the Andreev reflection probability has for $E=0$ declined by a factor of 10. This illustrates the high sensitivity to elastic scattering, which is important to superconducting hybrids. In Fig.~\ref{fig:Voss1984} one of the first comparisons with these theoretical predictions is shown using a molybdenum-tantalum point contact. With $Z$ as the only fit parameter  all the other curves are generated using Eq.\ref{Andreev_A}.        

Since the trajectory method did not have a cut-off for higher orders all contributions weighed equally and the sub-harmonic gap structure disappeared at lower temperatures. The obvious step to take was to introduce a finite transmissivity of the constriction, but the authors did not see a way to handle this. It led to an analysis of the much more simple problem of a ballistic S-c-N contact, which became the now well-known BTK paper\cite{BTK1982}. In a subsequent analysis we returned to the S-c-S case by putting two N-c-S contacts together~\cite{Octavio1983}, although we knew that it failed to describe the system properly in particular for low transmissivities of the double-barrier elastic potentials. While we were doing this work we received a copy in Russian of an article published by Zaitsev\cite{Zaitsev1980} in which he treated the S-c-S and the S-c-N case in the ballistic limit. It made us a bit nervous about the originality of our work. On the other hand he treated only a fully transmissive case, without any elastic scattering, which we felt was our most important innovation. Moreover he predicted an enhancement of the conduction by a factor of 3, which we saw as a sign that despite of the mathematical skills, the physical content was not fully appreciated (an erratum appeared soon)\cite{Zaitsev1980E}. Undoubtedly, the major step forward we made was the inclusion of elastic scattering, which became possible within the formalism that Zaitsev used, only after the introduction of new boundary conditions in 1984\cite{Zaitsev1984}. 
   
One of the very interesting aspects of the BTK-paper is that it made it possible by a simple point contact technique, pioneered by Sharvin and developed further by Yanson, Jansen \emph{et al.}, and Tsoi to read off from the derivative of the IV-curve the energy-dependence of the Andreev reflection coefficient as already appearing in the original paper by Andreev in 1964. I emphasize the energy-dependence, as was clearly shown by Blonder and Tinkham\cite{Blonder1983}, but was in retrospect already present in the data of Gubankov \emph{et al}~\cite{Gubankov1979}.  The first systematic use of this opportunity, shown in Fig.\ref{fig:Voss1984}, in which a set of conductance curves is given for a Mo-Ta point contact from the PhD Thesis of Gerhard Voss in Cologne (a student of Wohlleben). The quantitative success and the detailed dependence on the energy led to the emergence of Andreev point contact spectroscopy. Another important aspect is the ballistic nature of the assumed conduction of the contact, which is apparently justified despite the crude fabrication technology. The simplicity of the technique has made it possible to use it in a laboratory-course to train undergraduate students\cite{Luna2012}.  

\begin{figure}[t]
\includegraphics[width=7cm]{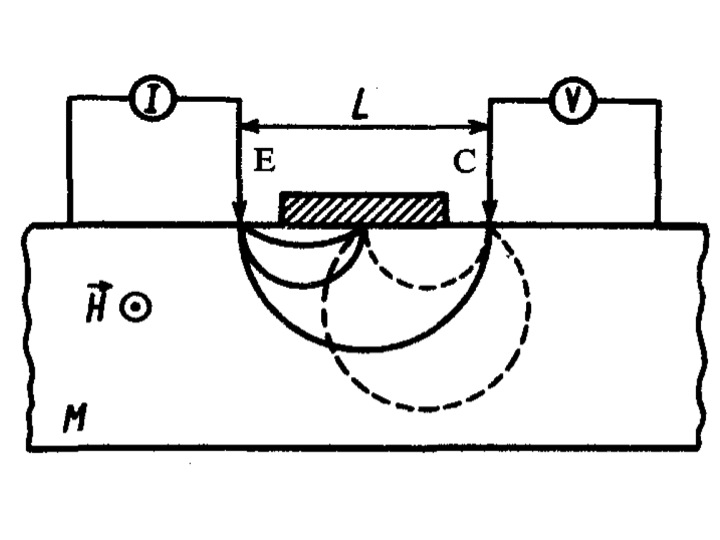}
\centering
\caption{\label{fig:RetroExp}A single crystal of bismuth is partially covered with a thin film of tin with a critical temperature of about about 3.8 K. Electrons injected at the point contact E reach the collector C following classical trajectories controlled by the magnetic field $\vec H$. Full curves electron-trajectories, dashed curves hole-trajectories after Andreev reflection below $T_c$. Taken from Bozhko \emph{et al.}\cite{Tsoi1982}.}
\end{figure}

\begin{figure}[t]
\includegraphics[width=8cm]{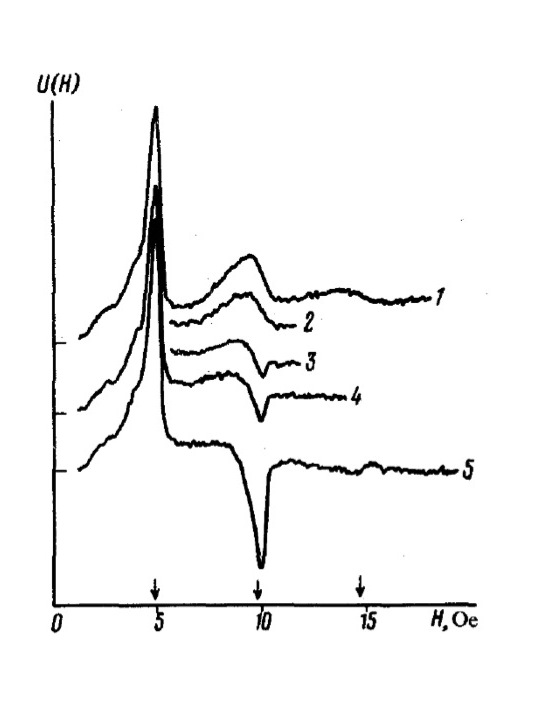}
\centering
\caption{\label{fig:RetroData}The electron-focusing signal observed in the collector contact for temperatures ranging from top to bottom of 3.80, 3.78, 3.74, 3.70 and 2.78 K. It clearly shows the emergence, at a field-strength of 10 Oersted, initially a positive contribution, which becomes  negative for lower temperatures. A very clear prove of both the (almost) identical, but reverse, momentum and the reversal of the charge. Taken from Bozhko \emph{et al.}\cite{Tsoi1982}.}
\end{figure}

The ballistic Sharvin-type point contacts have made it possible to measure directly the energy-dependence of the Andreev reflection coefficient, which has now been applied to a large variety of superconducting materials and the relevant theory has been generalized. Very recently, the Sharvin point contact idea was extended to apply to correlated materials as well by Lee et al\cite{LeePhillips2014}. Nevertheless, one of the hall-marks of Andreev reflection is the idea of retro-reflection. This is universally assumed to be one of the properties contained in the experimental data and sometimes explicitly assumed in the calculations. However, a direct demonstration of retroreflection itself is an interesting experimental challenge. The explicit demonstration started in the work Sharvin and Tsoi on point contacts with high-purity crystals. A  direct demonstration was carried out by Bozhko \emph{et al.}\cite{Tsoi1982} at Chernogolovka and Benistant \emph{et al.}\cite{Benistant1983} in Nijmegen using what is called electron focusing. It requires the ballistic transport from a point contact used as an emitter, the reflection from a superconductor, and the subsequent absorption by a second point contact which serves as a collector (Fig.\ref{fig:RetroExp}). Depending on the strength of the magnetic field the cyclotron orbits will coalesce at the absorbing point contacts. At specific values for the magnetic field, the lowest value given by $B_{focus}=2{\hbar k_F}/{2L}$ with L being the distance between the emitter and the collector a maximum signal is found. For increasing magnetic field two peaks will be found. For elastic scattering it will be two with the same sign. For Andreev reflection it will be a second one with an opposite sign because of the opposite charge. And if it is indeed retroreflection the hole trajectories should be copies of the electron trajectories.  This is exactly what is found and shown in Fig.\ref{fig:RetroData}. This last point contact experiment completed the demonstration of the essential ingredients of Andreev reflection: the energy-dependence, the charge reversal and the time-reversed paths.   
  
\section{Quantum transport in a point contact:~1988}

In the '80s the world of mechanically-made point contacts was gradually being transformed into a world of nanostructures made through clean-room technology.  In this transition the concept of Andreev reflection became fully embedded in the nanoworld. However, the first step was to take the idea of a Sharvin point contact from a classical concept with ballistic trajectories into the currently dominant paradigm of quantum transport. In the normal state the transport properties in small-scale structures are treated with the transmission-matrix formalism. Since the devices became small enough that quantum coherence was maintained over the size of the device transport became dominated by quantum interference. The emergence of this approach has greatly benefited from the increasing capability of making with lithography, in particular electron-beam lithography, metallic structures on a nanoscale.  Research objects from condensed-matter systems were made in which the length scale became a very important parameter. The research-objects are often called nano-devices or nano-structures, although in particular the name 'device' is somewhat misleading. They are usually not intended to provide real functionality, but rather to serve as a vehicle to reveal physical phenomena. In that sense it is part of the experimentalist's toolbox. For some the Large Hadron Collider in Geneva is the experimental system to do scattering experiments with, for others it is a nano-device of which many can be made in clean rooms worldwide. In combination with a superconductor a nanodevice is unavoidably an inhomogeneous superconducting system. It consists of different assemblies of atoms in which different states of matter can occur. The central concept to describe their transport is the scattering matrix, consisting of asymptotically free incoming states through an interaction region and providing free outgoing states. A recent review on the theoretical aspects is provided by Lesovik and Sadovskyy\cite{Lesovik2011}. It is  particularly specific about in what way it is justified in comparison to the more conventional kinetic equation and Green's function approaches and also what assumptions are being made which have to be satisfied in experimental systems.
 
\begin{figure}[t]
\includegraphics[width=8cm]{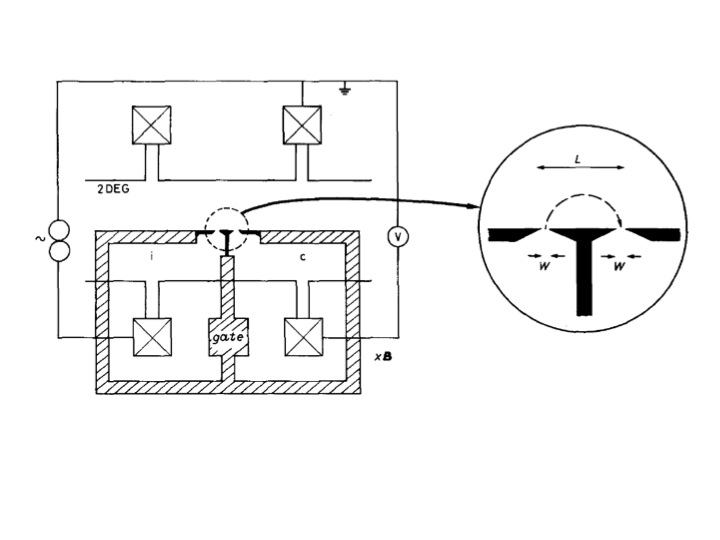}
\centering
\caption{\label{fig:VanHouten1}The experimental arangement of the gates on top of a GaAs/AlGaAs 2-dimensional electron gas. By changing the gate-voltage the 2DEG is depleted and transport is only possible in the gaps froming the point contacts. Given the low carrier density of 2DEG compared to a metal the transport was found to be quantized. The design was made to allow for an  electron-focusing experiment analogous to the one carried out by Tsoi \emph{et al.}\cite{Tsoi1974}. Figure taken from Van Houten \emph{et al.}\cite{Houten1988}.}
\end{figure}
The resistance in the normal state in the BTK result was clearly in the spirit of the early proposition of Landauer\cite{Landauer1970} about electrical conduction as a quantum transport phenomenon and given by: 
\begin{equation}
\label{Landauer}
G=\frac{e^2}{\pi\hbar}\frac{T}{R}
\end{equation}
with $G$ the conductance of a conductor and $T$ and $R$ the transmission and reflection coefficients.  The Landauer-formula, Eq.\ref{Landauer}, has been generalized to multi-channels by B\"uttiker \emph{et al.}\cite{Buttiker1985} and expressed as:

\begin{equation}
\label{MultichannelLandauer}
G=\frac{e^2}{\pi\hbar} \sum_{n,m=1}^{N_c} {|t_{nm}|}^2
\end{equation}

with $t_{nm}$ the transmission coefficient for scattering through the contact from incoming channel $n$ into outgoing channel $m$. The case with no elastic scattering in the wire, the perfect Sharvin contact, corresponds to $|t_{nm}|^2=\delta_{nm}$. This appealing scattering matrix approach was introduced to deal with normal transport in small structures with a length smaller than the phase-breaking length, over which phase-coherence was preserved.

At about the same time, the discovery of the Quantum Hall effect in 1980 by Von Klitzing et al\cite{Klitzing1980} in silicon MOSFETs, followed by the discovery of the Fractional Quantum Hall effect by Tsui et al\cite{FQHE1982} in GaAs/AlGaAs heterostructures, led to a strongly increased interest in 2-dimensional systems with a high mobility.  Prior to these systems the ballistic transport could be realized only in large single crystals of well-behaving metals. With the semiconductor-technology new artificially made systems became available and were continuously improved. These semiconductor heterostructures provided the 2-dimensional analog of the large single crystals of the past with the advantage of being a fully 2-dimensional systems of which also the carrier-density could be changed with a gate. 
\begin{figure}[t]
\includegraphics[width=10cm]{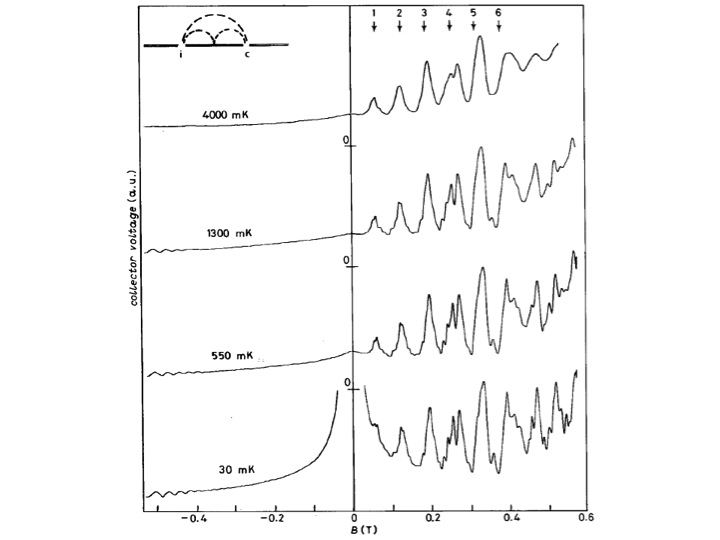}
\centering
\caption{\label{fig:VanHouten2}The electron-focusing signal observed in the collector-contact for a range of temperatures, clearly illustrating the ballistic nature of the transport. Due to the phase coherence fluctuations were observed reflecting the fact that the description has to go beyond the classical trajectories (and of course evolves into the Qauntum Hall effect). Figure taken from Van Houten \emph{et al.}\cite{Houten1988}. }
\end{figure}
These two developments, the microfabrication tools and the availability of new forms of matter in the form of heterostructures, came together in experiments carried out in a collaboration between Philips Research Laboratories and Delft University of Technology. Inspired by the point contact electron focusing experiments carried out by Tsoi et al\cite{Tsoi1974} and by Van Son et al\cite{VanSon1987} on single crystals of metals like silver, a two-point contact geometry was designed for a 2-dimensional electron gas in GaAs/AlGaAs (Fig.\ref{fig:VanHouten1}). The transport through one of these point contacts led to the surprising, but rapidly understood, discovery of quantized transport by Van Wees et al\cite{Wees1988} and submitted on Dec. 31, 1987. The electron-focusing experiments, exploiting the two contacts together (Fig.\ref{fig:VanHouten2}) were published separately by Van Houten et al\cite{Houten1988} and submitted, one week after the quantum point contact paper, on Jan. 6, 1988. This particular development towards the discovery of quantum transport, simultaneously with Wharam et al\cite{Wharam1988} inspired by a different conceptual tradition, shows nicely how the original idea of Sharvin on point contacts and of Tsoi on electron focusing had found its way to the modern lithography applied to condensed-matter structures, based on semiconducting heterostructures. Apparently, unaware of these recent developments Tsoi \emph{et al.}\cite{Tsoi1989} expected in a review in 1989 a development to use of electron focusing in the study of Andreev reflection in lithographically-made structures.

\section{Mechanical break-junctions: superconducting quantum point contacts:~ 1992}

The description of quantum transport with the transmission-matrix formalism has clearly a strong appeal also for superconductivity. It can easily be integrated with an analysis based on the Bogoliubov-De Gennes equations, assuming a ballistic transport system. An important problem is that a number of assumptions are being made, which are often difficult to meet in an experiment. As summarized clearly in B\"uttiker \emph{et al.}\cite{Buttiker1985} the model-system consists of 3 ingredients. There is a 'sample', which is characterized by a transmission matrix with the elements $t_{nm}$. There is no energy relaxation in the sample and transport through the sample is phase-coherent. The sample is connected on both sides to 'leads', which only serve to transport plane waves back and forth with probability $1$. Hence, they do not contribute to the scattering process. The leads are connected to 'reservoirs', which serve as equilibrium baths of electrons with a certain chemical potential and temperature. The value can be different in both reservoirs to represent an applied voltage-difference. 
\begin{figure}[t]
\includegraphics[width=8cm]{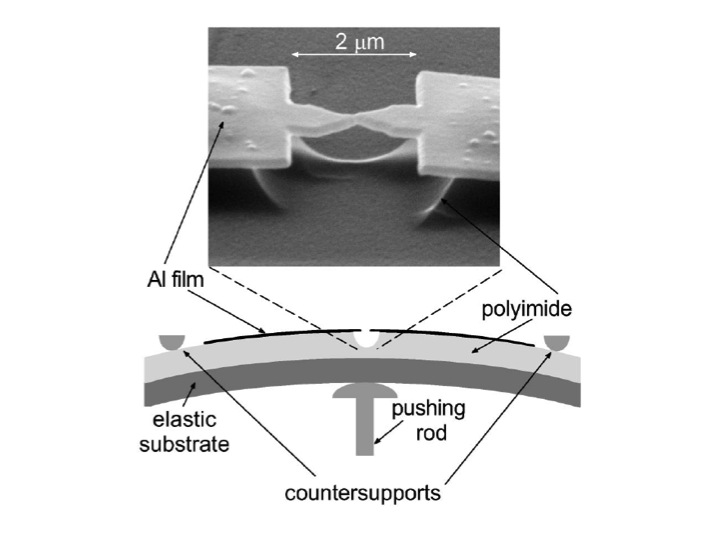}
\centering
\caption{\label{fig:ScheerMBJ} A controllable break junction configuration. It allows to gently break a wire, after which the two clean pieces can be brought together to form a vacuum tunnel contact, which evolves into a single atom point contact. Figure taken from Scheer \emph{et al.}\cite{Scheer1997}.}
\end{figure}
These assumptions are very reasonable for the conventional metallic point contacts. They are also valid for the quantum point contacts in a 2-dimensional electron gas, described in the previous Section. With the split-gate technique the major part of the 2 DEG is uneffected. Only at some point a constriction is created, whose width can be adjusted. The reservoirs are therefore of the same material as the constriction. The real physical contacts to the apparatus do not play a role, because the wide sections of the 2DEG serve as the equilibrium reservoirs.  The system is fully homogeneous in its properties, only the geometry of the conduction channel is changed. 

To achieve similar experimental conditions for a quantum transport channel with a superconductor is much harder. The unavoidable solution is that one has to combine two different materials. One that provides superconducting reservoirs and the other that acts as the 'sample'. These systems are therefore called 'superconducting hybrids' to which we will return below.  

\begin{figure}[t]
\includegraphics[width=8cm]{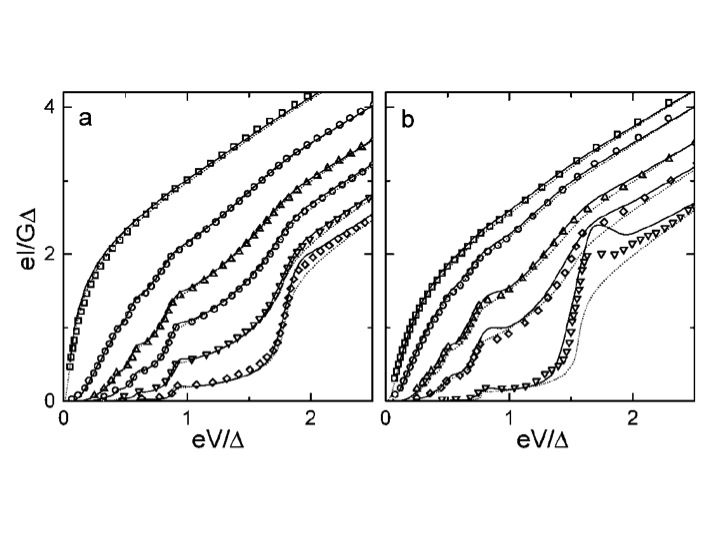}
\centering
\caption{\label{fig:Scheerprox}A set of IV-curves for increasing contact strength for Al (Panel a) and for Au, which is part of a bilayer of Al on top of Au (Panel b). Note the excellent agreement between theory and experiment in Panel a. In Panel b a clear demonstration is given of how multiple Andreev reflection is created by an induced pair-correlation in the Au, with some minor deviations accounted for by the theory. Figure taken from Scheer \emph{et al.} \cite{Scheer2001}.}
\end{figure}

The most natural link with the assumptions of the transmission matrix for superconducting nanotransport is that of the mechanical break junctions \cite{Muller1992} with an example shown in Fig.\ref{fig:ScheerMBJ}. By breaking a wire, which can then be gradually brought together again, single atom point contacts are created. The quantum conductor and the reservoirs are made of the same material in analogy to the GaAs/AlGaAs point contacts. In Fig.\ref{fig:Scheerprox} the left panel shows a set of I,V curves for single atom contacts of aluminium~\cite{Scheer1997, Scheer2001}. From bottom to top it shows the evolution of the IV-curve from weak contact to stronger contact. It is a modern version of the old Sharvin point contact with the clean room technology used to make a device which allows the breaking of a wire and the readjustment to bring the broken pieces together with subatomic precision. The data contain the same features observed in earlier generations of constriction-type Josephson jucntions. They include the phenomenon of multiple Andreev reflections or multiparticle tunneling leading to structure at voltages of $2\Delta/n$ and an excess current beyond a voltage of $2\Delta$.  In the right panel a similar set of curves is shown for a single atom Au-contact. However the Au is part of a bilayer  with Al. The Au has become superconducting through the proximity effect. The experiment very nicely illustrates how the Andreev reflections are coupled to the induced superconducting order in the Au. The differences can be largely accounted for by the standard diffusive proximity-effect theory using the quasiclassical equations. 

Since the tunneling strength is tunable by the delicate adjustment between the atoms both sets of curves can be understood as due to a limited number of conduction-channels, related to the orbitals of the aluminium or gold atoms. For increasing coupling strength, the transmission coefficients get closer to unity, the visibility of the structure weakens because higher-order processes are less damped. And since the material is the same there is no left-over mismatch between the two contacts limiting the transmission coefficients except for the orientation of the orbitals. This superconducting experimental system has the strong advantage that all experimental components of the quantum transport problem consist of one and the same material. 
  
\section{Superconducting heterohybrids: ~1985, 1990-1992}

The interest in superconductors in combination with semiconductors was initially not just driven by interest in quantum transport. It was also in response to the collapse of the superconducting Josephson-computer program.  From the mid-'60s the Josephson tunnel junctions had received wide-spread attention because of the program at IBM, Bell Labs, NIST, and various Japanese and European Laboratories to develop a digital Josephson computer, which would be fast and have a low power consumption. Around 1983 this highly visible Josephson-computer program of IBM was cancelled (earlier already at Bell Labs, and elsewhere it was quickly considerably reduced). In the aftermath it was extensively discussed that a Josephson-junction had an important drawback. It is a 2-terminal device, unlike the very successful semiconductor transistor, which is a 3-terminal device with gain: the output voltage, for example, can be larger than the input voltage. The gatability of a semiconductor as part of a superconducting device or some other scenario might help. 

Meanwhile many university laboratories with interest in superconducting thin films started to use lithography and advanced electron-beam lithography with modified SEM's. The interest in small area Josephson tunnel-junctions led at Bell Labs in 1977 to the very research-friendly Dolan-Dunkleberger\cite{Dolan1977, Dunkleberger1978} stencil lift-off technique or often called shadow-evaporation. Initially introduced as a technique for advanced photo-lithography it was, with the advent of electron-beam lithography, quickly used for dimensions in the hundreds of nanometers range. It allowed the development of thin-film devices consisting of multiple materials overlapping in certain areas with or without an oxide barrier in between. Therefore it was possible to combine superconductors with a normal metal, connected in multiple ways, and to measure, for example, the local density of states using Giaever tunneling. The same technique of shadow evaporation was also used to study phase coherent normal transport on a short length scale, related to the subject of weak localization, which arose at the end of the '70s, with the scaling theory of localization in 1979 as the famous hallmark.  On the fundamental side, the discovery of the Quantum Hall effect in 1980 and of quantized conductance  transport in 1988 led to the increased interest in high-mobility semiconductor heterostructures, in particular, if this could be combined with superconducting contacts. 

Although all these structures may be on a nanoscale I call them in this Section 'superconducting heterohybrids', because I am particularly interested in the path to ballistic transport, which appeared to be available in high-mobility heterostructures. The goal is to combine semiconductor heterostructures with superconducting contacts following a top-down technological path. Another kind of superconducting hybrids is based  on independently made nano-objects,  through a bottom-up process,  which can be further contacted with a superconductor (next Section). 

Electrical transport between a superconductor and a semiconductor is close to the subject of metal-semiconductor contacts. This has a long history, pretty much dominated by the subject of Schottky barriers. In this semiconductor-contact technology the best one can achieve is an 'ohmic' contact. It usually means that the IV-curve is linear, and physically it is the regime where the Schottky barrier is thin enough, by high doping, to have a current only due to quantum-mechanical tunneling through the Schottky barrier, without a thermally activated contribution. Assuming a degenerately doped semiconductor, acting like a metal, and the normal contact material in the superconducting state it would act as a normal metal-insulator-superconductor (NIS), a Giaever- like tunnel junction. However, given the resistance per unit area for these contacts the transmission probability $T$ is very low, in the order of  $10^{-4}$, and therefore Andreev reflection would not contribute significantly to the current. Moreover, for silicon with, for example, a material like lead (Pb) it was found that the Schottky barrier is not controlled by the work-function, but rather by details of the atomic ordering at the interface (see for example in Heslinga \emph{et al.}\cite{Heslinga1990}). In practice, contact formation is usually mixed with complex materials issues.  In the end it has so far been impossible to obtain interesting physics with superconducting contacts on silicon or with GaAs/AlGaAs heterostructures, which were so succesful for the Quantum Hall effect and the Quantum point contact. Instead the most successful results have been obtained with InAs-based semiconductors. This material is unique because the surface states lead to an inversion layer at the top of the crystal, which provides an easily accessible 2-dimensional electron gas. Currently, the interest in materials in which the surface acts as the conducting part has increased enormously. 

In the study of ballistic Andreev reflections the research with InAs-based heterostructures has provided at least one important experimental discovery.  In semiconductor-superconductor contacts, at relatively low temperatures compared to $T_c$,  a  zero-bias anomaly, compared to the canonical BTK-result,  was first reported by Kastalsky \emph{et al}~\cite{Kastalsky1991}. This anomaly consists of a peak in conductance centered around $V=0$, which increases upon lowering the temperature. In subsequent work it has become clear that it also occurs in normal metal-superconductor systems and is not unique to semiconductors. However, transport in semiconductors is closer to ballistic, which led Van Wees \emph{et al.}~\cite{VanWees1992} to explain it in terms of ballistic Andreev reflection modified by the impurity scattering in front of the $\delta -$function barrier introduced by Blonder \emph{et al.}~\cite{BTK1982} for elastic scattering. Since the single-particle phase is conserved it becomes possible that electrons are repeatedly scattered back, coherently, by the impurities to pass the  $\delta -$function barrier. It leads to the paradoxical behavior that adding impurity scattering \emph{enhances} the Andreev reflection probability. In other words, it corrects for the deletorious effects of the $Z-$parameter of Eq.\ref{E<}. 

In most of the remaining work on superconducting heterohybrids the focus was on the interplay between single particle phase-coherent transport, characteristic of the small scale of the normal metal, in interaction with the macroscopic phase of the superconductor. Since weak localization has a small effect and superconductivity a strong effect on the conductance, the dominating process is the effect of the superconductor on the normal metal which is known as the proximity effect. By using multiply connected devices many experiments were carried out, not possible before. An example is the study of phase-coherent normal transport controlled by scattering of electrons at different endings of a superconducting loop. In a SQUID-like fashion the conductance becomes dependent on the macroscopic phase difference controlled with a magnetic field applied to the loop. It leads to oscillations in the conductance of a normal metal wire, and it is often called Andreev interferometry, because it is understood as being due to phase-dependence of Andreev reflection. Some of these experiments have been reviewed in 2004 in Ref.\cite{Klapwijk2004}. Most of the experiments are in the diffusive limit and only assume the Andreev reflection as the process through which the information of the macroscopic phase is communicated.  The majority of these experiments can be interpreted with the diffusive quasiclassical non-equilibrium theory\cite{Belzig1999}. Consequently, they shed very little direct light on the ballistic Andreev reflection process itself. The advantage is however, that the experiments can be analyzed in many details in comparison to the well-developed diffusive theory. The best experimental system is a combination of a normal metal and a superconductor, rather than a semiconductor and a superconductor. A recent example is the work done recently by Vercruyssen \emph{et al.}~\cite{Vercruyssen2012} in which a superconducting nanowire was attached to two normal contacts at both ends. Instead of taking an old NS point contact configuration in the diffusive limit, a bulk N reservoir was connected to a superconducting wire, which was connected at the other end also to a large normal reservoir. This allows a one-dimensional analysis of the conversion of normal current to supercurrent, the evanescent states, another characteristic element of the Andreev reflection process taking place inside the superconductor, but in this case in the diffusive limit.  
 
\section{Superconducting nanohybrids:~1999-2002}

Since about 1999 the progress in creating nano-objects, usually through a chemical route or 'scotch tape',  has created a different type of superconducting nanostructures. They consist of bottom-up grown nano-objects, which can be found by inspection with an electron microscope, which also allows to contact them locally to electrical contacts.  Although normal contacts provide an interesting range of phenomena the use of superconductors as contacts contributes an extra energy and phase-coherence condition.  As expected, based on the universality of the Josephson effect, nano-objects used as a weak link between two superconductors carry a supercurrent with, depending on the geometry, some form of  Fraunhofer-like response to a magnetic field and, often,  the usual microwave-induced Shapiro steps. The experimentally and sometimes conceptually new aspect is that with a gate the nature of the nano-object and therefore the nature of the Josephson coupling can be tuned, which is also in competition with Coulomb interaction in these small structures. Since the coupling between the superconductor and the conductor plays a role, they are most often analyzed as quantum dots with superconducting leads. This has provided an interesting additional playground for model physics in which the effects of the spin (Kondo), the Coulomb blockade and the Josephson coupling can be explored\cite{SilvanoNN2010}. From an application point of view a gatable Josephson junction is potentially of interest too. It can act as a three-terminal device, like the field-effect transistor, a possibility which has been lacking for many years and has, as mentioned above, hampered the development of the digital Josephson computer. However, with the new approaches the gate-voltages used are in the tens of volts range, with an output voltage swing set by the $I_c R$-product, which is usually not much larger than a few tens of microvolts. So the gain of this type of transistor, a requirement for many practical applications, is absent. One of the few examples of a controllable Josephson junction, with an output voltage comparable to the input voltage is the superconducting transistor demonstrated by Morpurgo \emph{et al}~\cite{MorpurgoAPL1998}. This particular transistor or controllable Josephson junction allows the control of the Josephson current by controlling the occupation of states in the weak link, either by hot electrons or by a nonequilibrium distribution\cite{Baselmans1999,Baselmans2002}.    
\begin{figure}[t]
\includegraphics[width=8cm]{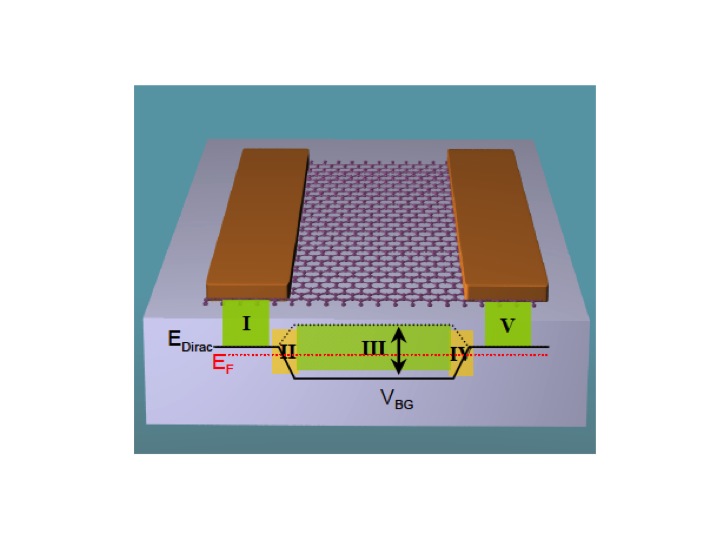}
\centering
\caption{\label{fig:Avouris}An indication of the various domains which contribute to the quantum transport for graphene coductors with normal metal electrodes.}
\end{figure}
The field of superconducting nanohybrids is currently a very active field with high expectations and with a multitude of theoretical proposals, in particular based on the new semiconducting nanowires. In the context of this overview it is premature to draw a conclusion about the experimental status. It may be more useful to indicate what the experimental difficulties are to come to robust experimental data when working with superconducting nanohybrids.  (Taken from Avouris \emph{et al.}\cite{AvourisNL2012}) 

\section{Experimental complexity of superconducting hybrids}

In the context of ballistic transport, superconducting heterohybrids and superconducting  nanohybrids are experimentally difficult to control. As pointed out above an attractive aspect of the GaAs/AlGaAs quantum point contacts and quantum dots in 2-dimensional electron gases and the mechanical break junctions is that they provide systems, which perfectly satisfy the requirements for the Landauer-B\"uttiker scattering apporach. The reservoirs are well defined and clearly distinct from the scattering region, because of their larger volume, while at the same time they are made of the same material. This is also true for the superconducting atomic point contacts. The hybrids are, by definition, built from different materials.  The reservoirs are made from a material which can go superconducting, to get a source and a collector of electrons in the superconducting state. Since this result is achieved through a multi-step clean room technology,  after fabrication of the device the first question to answer is what one has actually made. Usually, atoms forming the materials appear to be in the right spot, but this means usually not a lot from the point of view of the electrons. For semiconductor heterostructures, which are much more sensitive to dopants, the interfaces are made in UHV systems. For superconducting hybrids such an \emph{in situ} technology is not used and often not needed. Nevertheless, an important part of the experiments is the characterization of the device, usually by electrical transport, to determine what one has actually made. So, the characterization of the devices by electrical measurements gets intermixed with the identification of new physics and the choice of the best theoretical approach for the created nanostructure.  

In many experiments with superconducting hybrids a Josephson current is observed. However, a quantitative analysis turns out to be quite difficult. The observed critical current is usually smaller than expected, there is unaccounted for hysteresis observed in the IV-curves, there is often a lack of knowledge about the interface properties between the superconductor and the nano-object and finally, there is usually no quantitative analysis of the voltage-carrying state. In order to understand the transport processes in the superconducting hybrids better one needs a better understanding of the experimental system itself. The energy-dependent transport processes are the result of a mixture of elastic and Andreev reflection in a system in which ballistic and diffusive transport processes are inhomogeneously distributed.    

The same level of uncertainty occurs also with normal electrodes. An example was recently provided  by Avouris \emph{et al.}~\cite{AvourisNL2012}. In Fig.\ref{fig:Avouris} a pictorial summary is given of the regimes to be distinguished to understand short channel graphene-based quantum coherent ballistic transport\cite{AvourisNL2012}. It has metal films of 20 nm Pd with 30 nm Au on top of the graphene, with the graphene on $SiO_2$. Fabry-Perot resonances are observed for the electron branch of the IV-curve clearly signaling phase-coherent quantum transport. The length of the cavity is given by the uncovered part of the graphene, which indicates a reflection barrier at that interface, indicated by the Roman numeral II. In addition, a transmission coefficient between the metal and the graphene is identified, $T_{MG}$, which is specified to be in the order of 0.4. The origin of this transmission coefficient is more systematically studied in Ref.~\cite{Avouris2011NN}. In addition, they identify a gate-dependent transmission coefficient $T_K$ which is at the positions II. The main channel is area III, which is the channel which carries the Fabry-Perot resonances for the electrons. The absence of resonances for holes is attributed to the graphene underneath the metal being p-doped, meaning that the barrier in area II with $T_K$ is the result of a pn- and np-diode. Similarly, it is argued by Kretinin \emph{et al.}~\cite{Kretinin2010} for their InAs wires coupled to Al electrodes that the relevant length for the Fabry-Perot resonances is at the edge of the metal-covered part  and the uncovered part of the InAs. And finally, the same has been found experimentally in carbon nanotubes by Liang \emph{et al.}~\cite{Liang2001}. 
\begin{figure}[t]
\includegraphics[width=8cm]{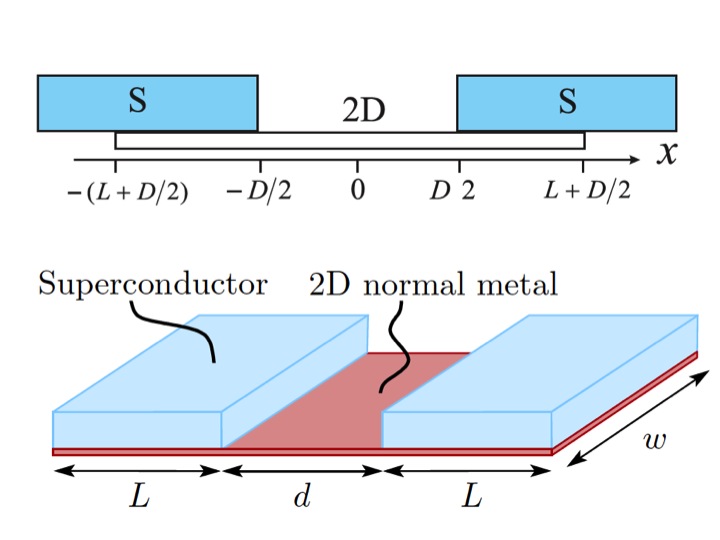}
\centering
\caption{\label{fig:kopnin}A picture taken from Kopnin \emph{et al.}\cite{Kopnin2014}showing the finite dimensions which contribute to the proximity-effect and therefore influence the energy-dependent conductance in a real device.}

\end{figure}

In most Josephson junction experiments a similar search for the limiting experimental parameters is needed. Since, most experiments focus on the gatability of the  Josephson current and a Josephson current can be established through a complex barrier of which the details are not urgently needed to know the details are not analyzed. An exception is in the work of Rohlfing \emph{et al.}\cite{Rohlfing2009}  in which for a Nb-InAs Josephson-contact for the transport channel a transmission coefficient, $T_{ch}$, of 0.8 is found, whereas for the interface between the 2DEG in the InAs and the superconducting metal a value, $\tau_{NS}$, of 0.06 is identified. Consequently a high number of multiple Andreev reflections signal a highly transmissive conduction channel, but does not signal a highly transparent superconductor-2DEG interface. 

At this point it is worth returning to the experiments by Scheer \emph{et al.}\cite{Scheer2001}. In this case an atomic scale contact of Au was used as the quantum-conductor coupled to bilayers of Al and Au. By studying the contacts in the tunneling limit they could measure the induced density of states in the N-part of the NS bilayer, which can be calculated using Usadel theory and which has extensively been measured with tunnel junctions, in fact it is the basis of the widely used niobium trilayer junction technology. So they have very accurate information about the induced proximity effect in N. Subsequently, they brought the Au electrodes together approaching the limit of highly transmissive channels. From a transmission-matrix point of view the reservoirs are now formed by the induced proximity effect in N, which does not display a standard BCS form with just a lower gap but has the well-known features such as a strong peak at the edge of the spectral gap. This induced superconducting state is now the source and drain for the transmission matrix, with a proximity-induced Andreev reflection coefficient, taking also into account that a certain length of the gold is not covered by the superconductor. The results are shown in Fig.\ref{fig:Scheerprox} (Panel b) with quite a good quantitative agreement, although not as good as for the full superconducting case in Panel a.      

Similarly, we can combine the insight obtained from the observations on the Fabry-Perot oscillations. They suggest that it is very plausible, depending on the details, that electron waves can also elastically scatter at the interface between the superconductor-covered and the uncovered part of graphene, a carbon nanotube or a semiconducting nanowire (Region III in Fig.\ref{fig:Avouris}). And it is well-known that a small amount of elastic scattering has a strong effect on the physical appearance of the IV-curve. Consequently, the problem of a quantitative and conceptual understanding needs 3 different transmission coefficients: $T_{NS}$ at the interface between the superconductor and the 'normal' metal, $T_E$ at the edge of the covered and the uncovered part, and $T_{ch}$ the transmission coefficient of the actual quantum conductor (assuming that it is possible to split the system into a number of well-identified parts). 

Finally, one more aspect of the problem is the superconductor itself. It is usually a thin film of finite length. Obviously, the metallic point contacts which have served the field of point contact spectroscopy very well are providing massive equilibrium reservoirs. In thin-film microbridges it was found important to make variable-thickness bridges in order to avoid thermal runaway at lower temperatures. These requirements were also needed in an experiment to study the 2-point resistance of a superconducting wire between normal electrodes\cite{Vercruyssen2012}. Apart from thermal equilibrium reservoirs a finite size superconductor is also contributing to the proximity effect (Fig. \ref{fig:kopnin}). This was addressed recently by Kopnin \emph{et al.}~\cite{Kopnin2011,Kopnin2014}.

 The process of Andreev reflection is indeed at the core of the transport properties of superconducting nanohybrids, but it takes quite some effort for an experimentalist to find out exactly how. This is the main reason that the direct observation of the unique features of Andreev reflection is much more difficult in the modern-day nanostructures than in the old-fashioned point contact technology and used in point contact spectroscopy.
\begin{figure}[t]
\includegraphics[width=8cm]{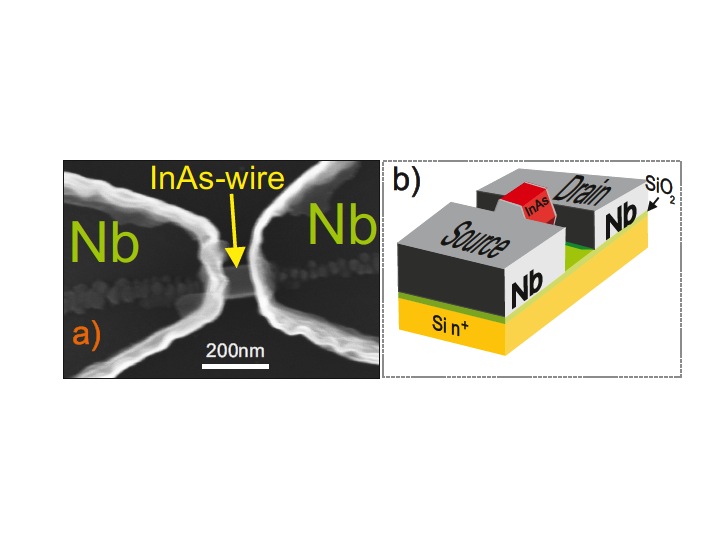}
\centering
\caption{\label{fig:InAskops}An experimental attempt to create a better defined contact between a superconductor and a semiconducting nanowired to reduce the number of unknown parameters and to create conditions for the electron-waves, which are susceptible to a theoretical analysis. Figure taken from G\"unel\cite{Gunel2013}.}
\end{figure}
Two fruitful approaches can be used. One of them is to invest quite a bit of experimental effort to disentangle all aspects of the problem. Along this path the recent paper by  Abay \emph{et al.}\cite{Abay2013} on InAs wires coupled to superconductors has made quite a bit of progress. The second approach is to simplify the experimental arrangement. An example is shown in Fig.\ref{fig:InAskops}, taken from G\"unel\cite{Gunel2013} A semiconductor nanowire is chopped off at the ends and the superconductor is attached at the tops. This should potentially reduce the number of relevant transmission coefficients. Similarly, such an approach has been used recently with a buried two-dimensional electron gas by etching a mesa and attaching the superconductor at the sides of the mesa, where it can reach the buried 2DEG\cite{Kononov2013}, which allows a combination of ballistic transport in the 2DEG together with superconducting contacts, analogous to the Tsoi experiments, but with the added option of the interaction with the Quantum Hall effect and/or Spin Hall effect. 

A final example,  is an approach in which new reservoirs are constructed with special properties as a means to discover new physics in conventional materials. An example of this approach is provided by Khaire \emph{et al.}~\cite{Birge2010}, to demonstrate the triplet proximity effect. The commonly used approach to the proximity effect is based on singlet Cooper pairs, which is the standard interpretation of the Josephson current in a SNS system for a diffusive system. For a thin ferromagnetic layer between two superconductors the Josephson coupling dies out rather quickly because of the large exchange energy in the ferromagnet, which leads to a very short coherence length. Khaire \emph{et al.} placed on both sides of the ferromagnet two modified reservoirs of the singlet superconductor. They modified, inspired by the work of Bergeret \emph{et al.}\cite{Bergeret2005}, the superconducting electrodes by covering them with a thin layer of normal metal followed by a very thin ferromagnet. This sandwich acted, through the thin ferromagnet, as a converter of singlet pairs into triplet pairs. They demonstrate that such triplet pairs have a long coherence length in a ferromagnet, in contrast to the singlet pairs. Although this experiment is performed in a diffusive system it illustrates very nicely that a creative modification of the reservoir can modify the transport through, in this case,  a ferromagnet considerably. 

These examples illustrate that the superconductng hybrids are very interesting and rich in potential. A lot more is to be expected, but they require very advanced materials control and extensive characterization, which takes time. As we will see time was also needed to measure directly the Andreev-bound states predicted in 1965.      

\section{Phase dependence and Andreev-bound states:~1969, 1992, 2013}

The last and very important characteristic of the Andreev reflection process is the dependence on the macroscopic quantum phase of the superconductor. The pair-potential $\Delta$ appearing in Eq.\ref{BDGeq} is a complex quantity with a well-defined phase $\phi$. It becomes very relevant when phase coherence in a normal metal close to a superconductor is measured or when two superconductors are coupled through the process of Andreev reflection. The phase dependence is not emphasized in the original paper by Andreev\cite{Andreev1964a}. It is also not very visibly present in a subsequent paper on the electronic states in the normal domains of a superconductor in the intermediate state. It is demonstrated that the energy levels are quantized\cite{Andreev1965}. This quantization is then used to calculate several thermodynamic quantities. In 1969 Kulik\cite{Kulik1969} addressed this phase dependence by pointing out that a bound state already assumes phase coherence, which if the two superconductors have a different phase makes the bound state energies dependent on the phase difference, leading to a discrete set of phase-dependent energies:     
\begin{equation}
\label{KulikBoundStates}
E_n^{\pm}=\frac{v_F}{2d}[2(\pi n+\alpha)\mp\phi]
\end{equation}
with $E_n$ the $n-$th energy level, $v_F$ the Fermi velocity, $d$ the thickness of the normal layer, $n$ an integer and $\phi$ the difference between the phases of the superconductors, $\phi_1-\phi_2$. The quantity $\alpha$ is weakly energy-dependent, which becomes more relevant when $E_n$ is closer to $\Delta_0$ the energy gap in the superconductor:    
\begin{equation}
\label{KulikBoundStates}
\alpha(E)=\arccos \big(\frac{E}{\Delta_0}\big).
\end{equation}
The levels $E_n$ are twofold degenerate for $\phi=0$ and split apart for $\phi\neq 0$, which is microscopically why there is a supercurrent running for a difference in the phases $\phi_1$ and $\phi_2$, and the reason that there is a net, Josephson current, for $\phi\neq 0$.  Kulik used this analysis to calculate the supercurrent in a SNS system, with a scattering free normal region. Beenakker and Van Houten~\cite{BeenakkerVanHouten1991} have used the same approach for a one-dimensional model in which the normal domain is short compared to the coherence length. In that case the thickness dependence disappears and the quantity $\alpha$ plays a key role leading to a single twofold-degenerate Andreev level:   
\begin{equation}
\label{ShortChannelAndreevClean}
E_A=\pm\Delta \cos(\phi/2)
\end{equation}
If there is scattering in the conduction channel with a transmission $\tau$ the Andreev levels are given by: 
\begin{equation}
\label{ShortChannelAndreev}
E_A=\pm\Delta\sqrt{(1-\tau\sin^2(\phi/2))}
\end{equation}
Ever since the first theoretical identification of discrete energy levels by Andreev and Kulik several attempts, have been made to measure directly these discrete levels by some form of spectroscopy. The lack of success until recently was partically because  in most experimental systems the requirement of one-dimensionality in a ballistic system was not fulfilled. A first indication was provided by Morpurgo \emph{et al.}~\cite{Morpurgo1997} by studying normal transport through a semi-ballistic coherent conductor of InAs of an InAs/AlSb heterostructure. The conductor was on both sides coupled to a superconductor connected in a loop. By applying a magnetic field the phase difference on both sides of the conductor could be tuned leading to the observation of a broad feature which was consistent with an Andreev bound state. For a diffusive system the supercurrent is not carried by a discrete set of states, but by a continuum of states. This continuum of states depends also on the phase-difference as has been shown very clearly by Baselmans \emph{et al}~\cite{Baselmans2002} in creating a so-called $\pi$-junction by selectively populating the states in the N-part of a SNS junction. A full experimental observation of the discrete Andreev levels has been achieved only very recently. Pillet \emph{et al.}~\cite{Pillet2010} studied carbon nanotubes connected to superconductors at both ends. By attaching a third electrode to the middle of the nanotube they were able to measure the individual states by quasi-particle injection spectroscopy. A very clear evolution of the individual states was observed, periodic in the phase-difference, as expected from Eq.\ref{ShortChannelAndreevClean}. A gate voltage was used to bring these quantum dot devices in the right regime for the observation of these Andreev bound states. Most recently, Bretheau \emph{et al}~\cite{Bretheau2013}, have used microwave spectroscopy in a mechanical break junction to probe the discrete levels and to occupy them selectively, taking into account the finite transmission coefficients of the conduction channels in the atomic scale point contact, as contained in Eq.\ref{ShortChannelAndreev}.  This work is, after many decades, the first experimental demonstration of the concept of phase-dependent Andreev bound states in a ballistic system, as first introduced by Kulik in 1969, and as a natural extension of the quantized levels introduced by Andreev in 1966.           

\section{Conclusions:~50 years later}
This review on the experimental proofs for the process called Andreev reflection leads to a somewhat surprising conclusion. The strongest evidence is provided by the experimental techniques based on point contacts developed prior to the modern era of nanotechnology. The more recent atomic-scale point contacts, which form a hybrid between the old-fashioned point contact technique and thin-film technology, have also contributed very significantly to the quantitative evaluation of several aspects of Andreev reflection. Such a quantitative evaluation has also been possible in many experiments based on diffusive inhomogeneous systems, in which the process of Andreev reflection is much more hidden and the theory is more complex. In the more recent semi-ballistic or partially ballistic thin-film-based superconducting hetero- and nano-hybrids the quantitative characterization of the electron transport is unfortunately less developed. And without the knowledge of the relevant experimental parameters it is also more difficult to identify the most appropriate theoretical framework to interpret the results and to provide a quantitative evaluation. This ongoing research will undoubtedly continue for another decade or more. However, at this point in time, 50 years later, we can safely conclude that it has been an impressive \emph{tour de force} to arrive at such an extremely fruitful concept as Andreev reflection inspired by the relatively 'murky' experimental basis of the thermal conductivity of Type-I superconductors in the intermediate state. 

\section*{Acknowledgments}

We like to thank A.V.Semenov, F.S.Bergeret, S.N.Artemenko and 2 anonymous referees for a critical reading of the manuscript and for providing helpful comments for improvement. Financial support from the Ministry of Education and Science of the Russian Federation under Contract No. 14.B25.31.0007 and from the European Research Council Advanced grant  No. 339306 (METIQUM) is gratefully acknowledged.


\end{document}